\documentclass[preprint2]{aastex}
\usepackage{natbib}

\shorttitle{Fluorine in the solar neighborhood}
\shortauthors{J\¨onsson et al.}

\begin{document}

\title{Fluorine in the solar neighborhood \\ - is it all produced in AGB-stars?}

\author{H. J\"onsson and N. Ryde}
\affil{Lund Observatory, Department of Astronomy and Theoretical Physics, \\ Lund University, Box 43, SE-221 00 Lund, Sweden}
\email{henrikj@astro.lu.se}

\author{G. M. Harper}
\affil{School of Physics, Trinity College, Dublin 2, Ireland}

\author{M. J. Richter}
\affil{Physics Department, University of California, Davis, CA 95616, USA}

\and

\author{K. H. Hinkle}
\affil{National Optical Astronomy Observatory, P.O. Box 26732, Tucson, AZ 85726, USA}

\begin{abstract}
The origin of `cosmic' fluorine is uncertain, but there are three proposed production sites/mechanisms: AGB stars, $\nu$ nucleosynthesis in Type II supernovae, and/or the winds of Wolf-Rayet stars. The relative importance of these production sites has not been established even for the solar neighborhood, leading to uncertainties in stellar evolution models of these stars as well as uncertainties in the chemical evolution models of stellar populations.

We determine the fluorine and oxygen abundances in seven bright, nearby giants with well-determined stellar parameters. We use the 2.3 $\mu$m vibrational-rotational HF line and explore a pure rotational HF line at 12.2 $\mu$m. The latter has never been used before for an abundance analysis. To be able to do this we have calculated a line list for pure rotational HF lines. We find that the abundances derived from the two diagnostics agree.

Our derived abundances are well reproduced by chemical evolution models \emph{only} including fluorine production in AGB-stars and therefore we draw the conclusion that this might be the main production site of fluorine in the solar neighborhood. Furthermore, we highlight the advantages of using the 12 $\mu$m HF lines to determine the possible contribution of the $\nu$-process to the fluorine budget at low metallicities where the difference between models including and excluding this process is dramatic.

\end{abstract}

\keywords{stars: abundances --- solar neighborhood --- molecular data}

\section{Introduction} \label{sec:intro}
`Cosmic' production of fluorine is difficult because fluorine is very easily destroyed in stellar interiors and therefore has to be deposited into the interstellar medium soon after its production. Because of this sensitivity to the conditions of its production site, the `cosmic' fluorine abundance will not only put a severe constraint on the chemical evolution models describing different stellar populations, but also on stellar evolution models.

Three production sites/mechanisms have been proposed to contribute to the `cosmic' fluorine abundance: thermal-pulsing asymptotic giant branch (TP-AGB) stars, $\nu$ nucleosynthesis in supernovae type II (SNeII), and/or Wolf-Rayet (W-R) stars (see \citet{2014A&A...564A.122J} for further details).
So far, only the production of fluorine in AGB-stars has been proven by observations: by direct measurements of fluorine abundance in AGB-stars \citep{1992A&A...261..164J,2009ApJ...694..971A,2010ApJ...715L..94A}, by measurements of fluorine in post AGB-stars and planetary nebulae \citep{2005A&A...433..641W,2005ApJ...631L..61Z,2008ApJ...682L.105O} as well as in carbon-enhanced metal-poor stars \citep{2007ApJ...667L..81S,2011ApJ...729...40L}, and in Ba-stars \citep{2011A&A...536A..40A}. Also fluorine pollution by AGB-stars in globular clusters has been shown by for example \citet{2013ApJ...763...22D}. When it comes to the production of fluorine by the $\nu$-process, \citet{2005ApJ...619..884F} do not see any evidence for it in the interstellar medium and the fluorine production of W-R stars has been theoretically questioned by \citet{2005A&A...443..243P}. This means that these two production sites are more speculative at the moment.

To determine the relative role of the three production processes more observations are needed. However, determining the fluorine abundance is not easy because of a lack of spectral lines in stellar spectra. The HF line at 2.3 $\mu$m is often used, but it is very weak in dwarfs and metal-poor giants. Unfortunately it is situated in a region with a lot of telluric lines adding even more uncertainty to the fluorine abundance determined from this line \citep{2013A&A...560A..74D}. Furthermore there have been several sets of molecular data for the HF molecule used in the literature: one from \citet{1992A&A...261..164J}, in turn from Tipping and one from \citet{2000PhDT........16D}, in turn from Sauval, differing in the excitation energy by 0.25 eV.
For further details on these line lists, see \citet{2014A&A...564A.122J}, where a HF line list compatible with the partition function built into many spectral synthesis programs (for example MOOG, BSYN, and SME) was presented. Shortly thereafter \citet{2014arXiv1404.5755M} published a version of the HF line list based on Einstein $A$-values from the HITRAN2012 database. These $A$-values are very close to the values used in \citet{2014A&A...564A.122J} leading to an agreement of the log$gf$ values within $\sim$0.01 dex.

When it comes to fluorine production in the solar neighborhood, as mentioned earlier, \citet{1992A&A...261..164J} and \citet{2009ApJ...694..971A,2010ApJ...715L..94A} showed production in AGB-stars. \citet{2012A&A...538A.117R} argued that the main production site is AGB-stars, while \citet{2013AJ....146..153N} claim that the relative fluorine contribution from AGB-stars probably are \emph{not} the main source of fluorine in the solar neighborhood, but rather the $\nu$-process. In line with the modeling of \citet{2011ApJ...739L..57K}, who predict that the largest difference in fluorine abundance for a scenario with and without the $\nu$-process can be found in metal-poor stars, \citet{2013ApJ...765...51L} explore  metal-poor field giants, and show that neither model fit their observations well. However, the model closest to the observed values is the one including the $\nu$-process.

Obviously the question of the fluorine abundance trend in the solar neighborhood is still open, and there is indeed need for further study.

In this \textit{Letter}, we reevaluate the chemical evolution of fluorine in the solar neighborhood by comparing current models with newly derived fluorine abundances for seven bright giants using our new line lists, including the HF lines around 12 $\mu$m, which are here used for the first time. The 12.2 $\mu$m line used is much stronger than the 2.3 $\mu$m line and is not affected by telluric lines.

\section{Observations}
Spectra of seven bright, nearby giants in the 12.2 $\mu$m region were recorded with the spectrometer TEXES \citep{2002PASP..114..153L} mounted on IRTF on Mauna Kea during 2000, 2001 and 2006. The spectra were extracted and reduced in a typical manner (see \citet{2002PASP..114..153L} for further details) and have a resolution of $R\sim$65,000 and a signal-to-noise ratio of typically 100.  Furthermore, we have retrieved four spectra covering the 2.3 $\mu$m HF line observed with the FTS mounted on the Kitt Peak National Observatory Mayall 4 m reflector (one of the four spectra is the IR Arcturus atlas by \citet{1995iaas.book.....H}). These observations were made on June, 24 1977, August, 24 1983, and on April, 13 1990.
For the determinations of metallicity and oxygen abundances, we searched spectral archives for visual spectra of our target stars and found three from the NARVAL spectrometer, one from HARPS \citep{2003Msngr.114...20M}, and one from ELODIE \citep{1996A&AS..119..373B}.

\section{Analysis}
All spectra were analyzed using the software \texttt{Spectroscopy Made Easy}, SME \citep{1996A&AS..118..595V} and a grid of MARCS spherical symmetric LTE models \citep{2008A&A...486..951G}. 

\subsection{Stellar parameters}\label{sec:params}
The stellar parameters used are listed in Table \ref{tab:stars}. Effective temperatures determined from angular diameter measurements are taken from \citet{2003AJ....126.2502M}, and gravities are determined from the stellar radius \citep{2003AJ....126.2502M}, the parallax \citep{2007A&A...474..653V}, and fits to evolutionary tracks (see Ryde et. al (in prep.) for further description). The [Fe/H] and the optical microturbulence were, for five of our stars, determined from Fe I-lines in visual spectra from the three different spectroscopical archives. For the remaining two stars we could not find any optical spectra, but use literature values for the metallicity, see Table \ref{tab:stars}. For the IR spectra we, as suggested by \citet{2008A&A...489.1271T}, instead use a slightly higher value of 2.0 km/s for the microturbulence for all stars.

Typical uncertainties of the stellar parameters are $\sigma T_{\mathrm{eff}}$=50 K, $\sigma \log g$=0.1, $\sigma$[Fe/H]=0.1, and $\sigma v_{\mathrm{mic}}$=0.5 kms$^{-1}$. 

\begin{deluxetable}{c r c c c c c c c c c c}
\tablecaption{Stellar parameters and abundances for our program stars.\label{tab:stars}}
\tablewidth{0pt}
\tablehead{
\colhead{Star} & \colhead{HD} & \colhead{$T_{\mathrm{eff}}$} & \colhead{logg} & \colhead{[Fe/H]\tablenotemark{a}} & \colhead{$v_{\mathrm{mic}}$\tablenotemark{b}}& \colhead{$v_{\mathrm{mic}}$\tablenotemark{c}} & \colhead{$A$(O)$_{[\mathrm{O I}]}$} & \colhead{$A$(O)$_{\mathrm{OH}}$} & \colhead{$A$(O)$_{\mathrm{mean}}$} & \colhead{$A$(F)$_{2.3\mu}$} & \colhead{$A$(F)$_{12.2\mu}$}}
\startdata
$\delta$Vir & 112300 & 3602 & 0.84 & -0.14\tablenotemark{2} & \nodata & 2.00  & \nodata               &    8.65 &    8.65 & \nodata & 4.20 \\
$\delta$Oph & 146051 & 3721 & 1.02 & -0.24\tablenotemark{e} & 1.48\tablenotemark{e}  & 2.00  & 8.48\tablenotemark{e} &    8.53 &    8.50 &    4.02 & 4.04 \\
   $\mu$UMa &  89758 & 3793 & 1.07 & -0.34\tablenotemark{f} & 1.66\tablenotemark{f}  & 2.00  & 8.47\tablenotemark{f} &    8.43 &    8.45 &    3.94 & 4.05 \\
$\alpha$Lyn &  80493 & 3836 & 0.98 & -0.31\tablenotemark{1} & \nodata & 2.00  & \nodata               & \nodata & \nodata & \nodata & 3.96 \\
$\alpha$Tau &  29139 & 3871 & 1.27 & -0.25\tablenotemark{d} & 1.51\tablenotemark{d}  & 2.00  & 8.57\tablenotemark{d} &    8.53 &    8.55 &    4.16 & 4.34 \\
$\alpha$Hya &  81797 & 4060 & 1.35 & -0.17\tablenotemark{d} & 1.84\tablenotemark{d}  & 2.00  & 8.62\tablenotemark{d} &    8.54 &    8.58 & \nodata & 4.12 \\
$\alpha$Boo & 124897 & 4226 & 1.67 & -0.62\tablenotemark{d} & 1.65\tablenotemark{d}  & 2.00  & 8.60\tablenotemark{d} &    8.40 &    8.50 &    3.65 & 3.73 \\
\enddata
\tablenotetext{a}{We use A(Fe)$_{\odot}$=7.50 \citep{2009ARA&A..47..481A}.}
\tablenotetext{b}{microturbulence used for the visual spectra.}
\tablenotetext{c}{microturbulence used for the IR spectra.}
\tablenotetext{d}{As determined from NARVAL archive spectrum.}
\tablenotetext{e}{As determined from FEROS archive spectrum.}
\tablenotetext{f}{As determined from ELODIE archive spectrum.}
\tablerefs{(1) \citet{1990ApJS...74.1075M}; (2) \citet{1985ApJ...294..326S}}
\end{deluxetable}

\subsection{Line data}\label{sec:params}
The line data for all lines used are the same as in \citet{2014A&A...564A.122J}, except for the HF line at 12.2 $\mu$m. 
We use the partition function already presented in \citet{2014A&A...564A.122J}, which is an updated version of the one from \citet{1984ApJS...56..193S}. Since Equation 3 in  \citet{2014A&A...564A.122J} is not entirely correct, we give it here again, see below. The Figure 2 shown in  \citet{2014A&A...564A.122J} is however correct.  The partition function is given by
$\ln Q = \sum_{i=0}^{5} a_{i}\times (\ln T\mathrm{[K]})^i$ where\\

$\mathbf{a} = \left(\begin{array}{c}
-360.544650\\
222.384130\\
 -54.5664753\\
 6.69351087\\
-0.409637436\\
0.0100497602\\
 \end{array} \right)\label{part}$

The excitation energies were computed from the energy-level expression and coefficients of \citet{1994JMoSp.164..574L}. The transition frequencies were calculated from the differences of the energy levels involved in the transition, and agree excellently with accurately measured frequencies from \citet{1987JMoSp.122..477J}. The HF Einstein decay coefficients, $A_{ji}$, for the rotational transitions were computed using the accurate dipole moment found by \citet{1970JChPh..52.6033M}. The oscillator strengths, the $gf$ values, were then calculated with the conversion given in \citet{1983A&A...128..291L}. We used a statistical weight following our partition function, $g=2J+1$, where $J$ is the rotational quantum number. Our calculated data for the rotational HF lines are listed in Table \ref{tab:linedata}.

\begin{deluxetable}{ r r r r r c r c }
\tablecaption{HF rotational transitions$^{\mathrm{a}}$.\label{tab:linedata}}
\tablewidth{0pt}
\tablehead{
\colhead{$J'$} & \colhead{$J''$} & \colhead{$\sigma$} & \colhead{$\lambda_\mathrm{air}$} & \colhead{$\chi_{exc,low}$} & \colhead{$\chi_{exc,low}$} & \colhead{$A_{J',J''}$} & \colhead{$\log gf$}\\
& & \colhead{[cm$^{-1}$]} & \colhead{[{\AA}]} &  \colhead{[cm$^{-1}$]} & \colhead{[eV]} &  \colhead{[s$^{-1}$]}}
\startdata
   1  &  0  &    41.111 & 2431777.1601  &     0.00 & 0.000  &    0.024  &   -4.191\\
   2  &  1  &    82.171 & 1216640.9581  &    41.11 & 0.005  &    0.232  &   -3.589\\
   3  &  2  &   123.130 &  811930.5537  &   123.28 & 0.015  &    0.837  &   -3.237\\
   4  &  3  &   163.936 &  609827.2819  &   246.41 & 0.031  &    2.049  &   -2.988\\
   5  &  4  &   204.540 &  488767.6306  &   410.35 & 0.051  &    4.070  &   -2.795\\
   6  &  5  &   244.893 &  408230.6151  &   614.89 & 0.076  &    7.092  &   -2.637\\
   7  &  6  &   284.944 &  350850.2607  &   859.78 & 0.107  &   11.296  &   -2.505\\
   8  &  7  &   324.646 &  307943.7237  &  1144.73 & 0.142  &   16.847  &   -2.390\\
   9  &  8  &   363.951 &  274687.3322  &  1469.37 & 0.182  &   23.893  &   -2.289\\
  10  &  9  &   402.812 &  248187.0031  &  1833.32 & 0.227  &   32.564  &   -2.199\\
  11  & 10  &   441.184 &  226601.1585  &  2236.14 & 0.277  &   42.971  &   -2.118\\
  12  & 11  &   479.021 &  208702.1853  &  2677.32 & 0.332  &   55.203  &   -2.045\\
  13  & 12  &   516.281 &  193640.2976  &  3156.34 & 0.391  &   69.325  &   -1.978\\
  14  & 13  &   552.920 &  180808.6018  &  3672.62 & 0.455  &   85.383  &   -1.916\\
  15  & 14  &   588.899 &  169762.1355  &  4225.54 & 0.524  &  103.397  &   -1.858\\
  16  & 15  &   624.177 &  160167.2678  &  4814.44 & 0.597  &  123.363  &   -1.805\\
  17  & 16  &   658.717 &  151768.9583  &  5438.62 & 0.674  &  145.254  &   -1.755\\
  18  & 17  &   692.481 &  144368.9299  &  6097.33 & 0.756  &  169.024  &   -1.709\\
  19  & 18  &   725.435 &  137810.7354  &  6789.81 & 0.842  &  194.598  &   -1.665\\
  20  & 19  &   757.545 &  131969.3039  &  7515.25 & 0.932  &  221.884  &   -1.624\\
  21  & 20  &   788.780 &  126743.4749  &  8272.80 & 1.026  &  250.769  &   -1.585\\
  22  & 21  &   819.109 &  122050.5681  &  9061.58 & 1.123  &  281.119  &   -1.549\\
  23  & 22  &   848.504 &  117822.3714  &  9880.69 & 1.225  &  312.785  &   -1.514\\
  24  & 23  &   876.938 &  114002.1308  & 10729.19 & 1.330  &  345.601  &   -1.481\\
  25  & 24  &   904.385 &  110542.2649  & 11606.13 & 1.439  &  379.388  &   -1.450\\
\enddata
\tablenotetext{a}{The consistent partition function is given in the text.}
\end{deluxetable}

\subsection{Stellar abundances}
The iron abundances were determined from optical Fe I-lines or taken from literature sources (see Section \ref{sec:params} and Table \ref{tab:stars}). The oxygen abundances were determined from the FTS spectra using OH-lines around 1.56 $\mu$m, and when optical archive spectra were available, also from  the [O I] line at 6300 \AA. The final oxygen abundance used in Figures \ref{fig:feh}-\ref{fig:koba} is the mean value of these two. For four of our stars we have FTS K-band spectra where the 2.3 $\mu$m HF line is unaffected by telluric lines, so for that subset of stars we are able to compare the fluorine abundances as derived from the 12.2 $\mu$m HF line to the abundances from the 2.3 $\mu$m line. To our knowledge this is the first determination of fluorine for all stars presented here except $\alpha$Boo, which has been extensively studied because of the available atlas of \citet{1995iaas.book.....H}. For example \citet{2013AJ....146..153N} get a value of $A$(F)$=3.75$ from the 2.3 $\mu$m line, which is close to our value\footnote{\citet{2013AJ....146..153N} use the line list of Sauval, which is very similar to to our, and MOOG, which is distributed with a partition function compatible with these lists, so our abundance results most likely are on the same scale.}.

The uncertainties in the determined abundances from the uncertainties in the stellar parameters (see Section \ref{sec:params})  are generally small with $\sigma A$(O)$=0.1$, $\sigma A$(F)$_{2.3\mu}=0.1$, and $\sigma A$(F)$_{12.2\mu}=0.2$.
The 2.3 $\mu$m line is more temperature sensitive than the 12.2 $\mu$m HF line and, for our sample of stars, the 12.2 $\mu$m HF line is typically very sensitive to the microturbulence, the reason being that it is on the verge of being saturated for most of our stars ($\log W_\lambda / \lambda \geq -5.3$).

\begin{figure*}
\epsscale{2.00}
\plotone{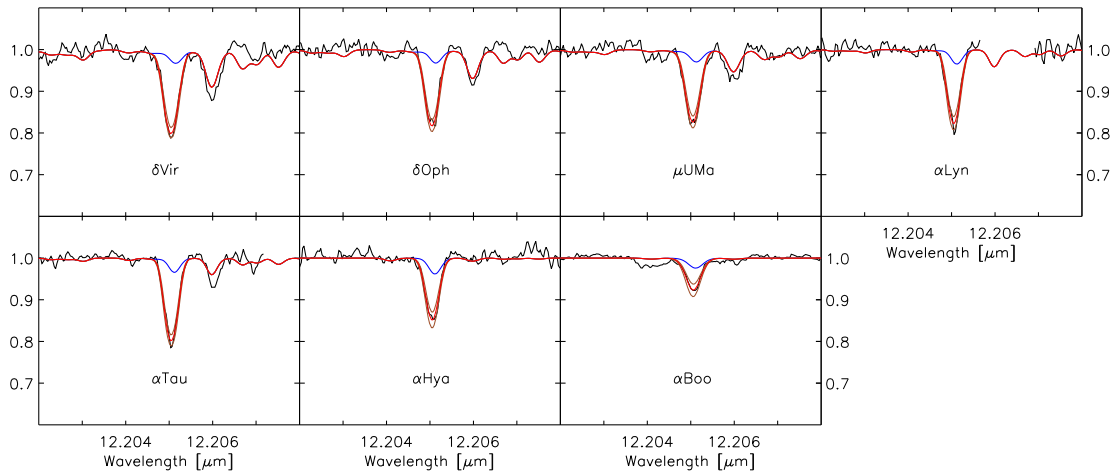}
\caption{Spectra showing the 12.2 $\mu$m HF line for our stars. Our best fit is shown in red and $\pm$0.2 dex is shown in brown. Note that the best fit is determined by $\chi^2$-minimization and simply comparing the synthetic spectra to the observed by eye would most probably lead to higher abundances for $\alpha$Lyn and $\alpha$Tau. The contribution of the blending Mg I line is shown in blue. All other lines are due to water and we note that some of them are not well reproduced in the synthetic spectra of the coolest stars. This will be explored in a future paper (Ryde et al. in prep).\label{fig:spectra}}
\end{figure*}

In Figure \ref{fig:spectra} we show the spectra around the 12.2 $\mu$m HF line for our stars. We note that the HF-line is blended with a Mg I line (122051.12 \AA, $\chi_{exc,low}=7.092$, and $\log gf=0.353$), but assuming the atomic data is correct (it is rated `B+' in the NIST database, meaning an uncertainty in the transition probability of $\leq 7\%$) SME will compensate for this line in the spectral fitting and the fluorine abundance determination. Several water lines are also present in the 12 $\mu$m region, which will be explored in a forthcoming paper (Ryde et al. in prep). 

\section{Results and discussion}
\begin{figure*}
\epsscale{2.00}
\plotone{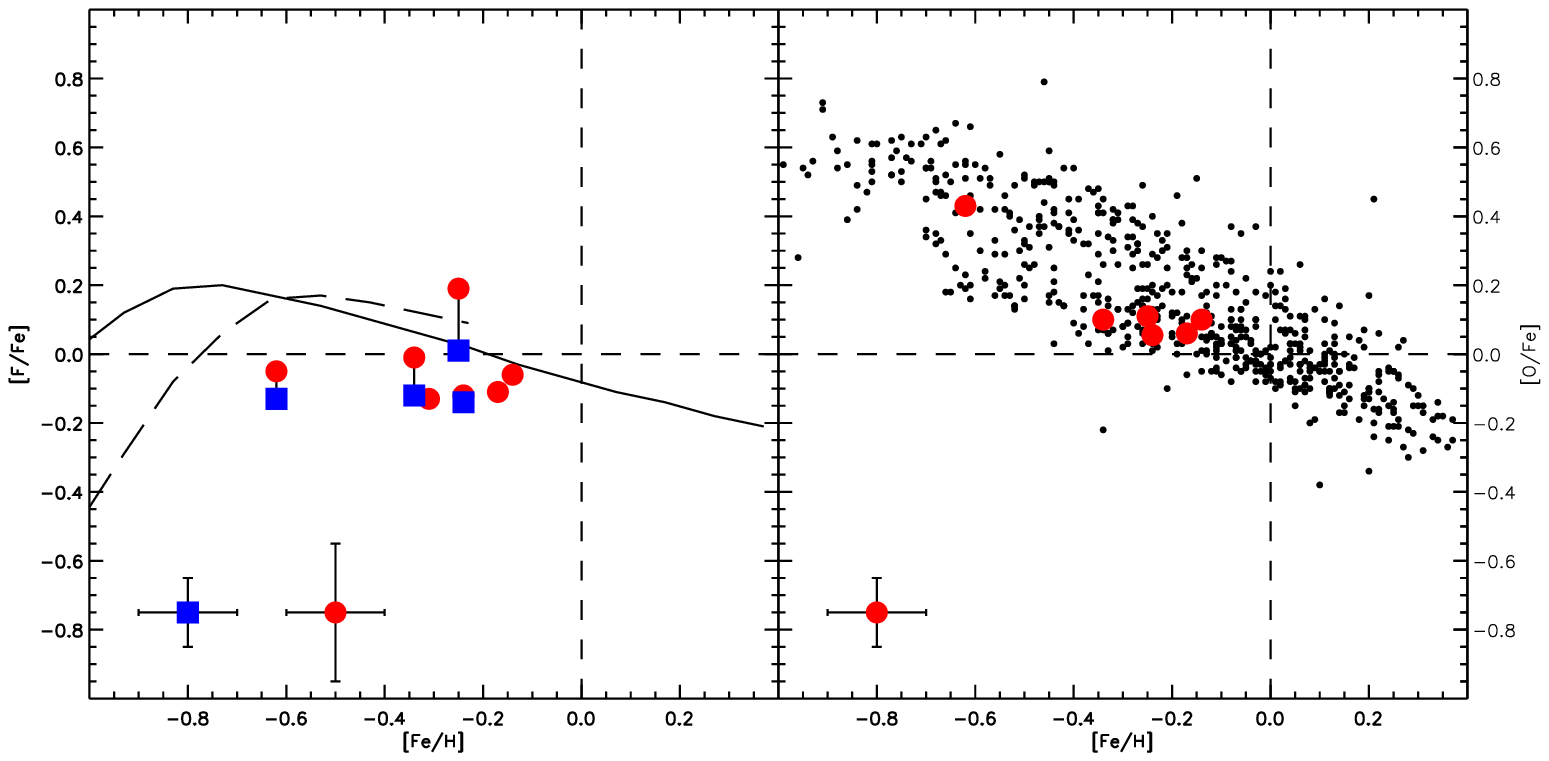}
\caption{[F/Fe] and [O/Fe] as functions of [Fe/H] for our program stars. Left panel: results from the 2.3 $\mu$m HF line are marked with blue squares and results from the 12.2 $\mu$m line are marked with red dots. Results for the same stars are interlinked with lines. Also shown are the predictions of the models from \citet{2011MNRAS.414.3231K} not including fluorine production in W-R stars and via the $\nu$-process. The full line is the solar neighborhood model and the dashed line is the thick disk model. The model predictions have been transformed to the solar abundance scale of A(F)$_{\odot}$=4.40 \citep{2014arXiv1404.5755M} and A(Fe)$_{\odot}$=7.50 \citep{2009ARA&A..47..481A}. Right panel: the oxygen abundances plotted are the mean of the abundances derived from the 6300 {\AA} [O I]-line and 1.55 $\mu$m OH-lines. The black dots are the solar neighborhood dwarfs of \citet{2014A&A...562A..71B} consisting of thin- and thick disk type stars showing the typical bi-modality of lower and higher oxygen enhancement, respectively. Conservative estimates of the uncertainties are marked in the lower left corners in both panels.\label{fig:feh}}
\end{figure*}

\begin{figure}
\epsscale{1.05}
\plotone{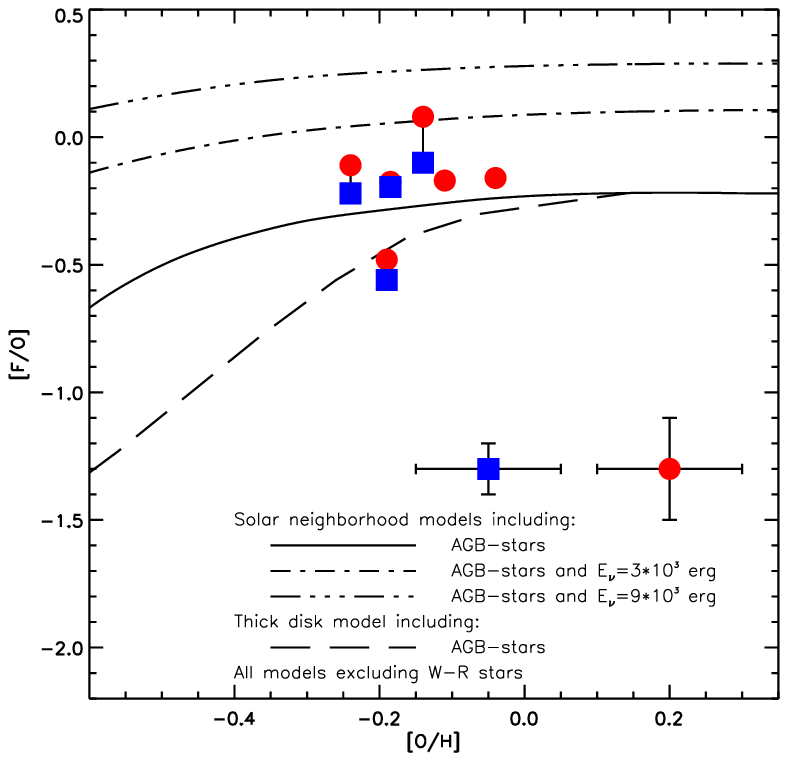}
\caption{Our fluorine abundances compared with the predictions of the models from \citet{2011ApJ...739L..57K} and \citet{2011MNRAS.414.3231K}. The model predictions have been transformed to the solar abundance scale of A(F)$_{\odot}$=4.40 \citep{2014arXiv1404.5755M} and A(O)$_{\odot}$=8.69 \citep{2009ARA&A..47..481A}. Results from the 2.3 $\mu$m HF line are marked with blue squares and results from the 12.2 $\mu$m line are marked with red dots. Results for the same stars are interlinked with lines. Conservative estimates of the uncertainties are marked above the text in the plot.\label{fig:koba}}
\end{figure}

Our abundance results are listed in Table \ref{tab:stars} and plotted in Figures \ref{fig:feh}-\ref{fig:koba}. The fluorine abundances as derived from the 2.3 $\mu$m line and the 12.2 $\mu$m line are in close agreement for all stars except $\alpha$Tau where the latter value is $\sim$0.2 dex higher than the former. However, because of the strong microturbulence dependence of the 12.2 $\mu$m line this is within the uncertainties. We note that changing the IR microturbulence, within the uncertainty, to 2.5 km/s will shift this value into the trends of the others in the plots.

Since the fluorine abundances derived from the 2.3 $\mu$m and the 12.2 $\mu$m lines agree so well using standard MARCS atmospheres for these red giants, we can conclude that also the formation of the 12 $\mu$m lines are well described by such
models. On the contrary, numerous water lines in this wavelength region are poorly modelled. \citet{2002ApJ...580..447R,2006ApJ...637.1040R} thus constructed a semi-empirical model atmosphere which could explain the formation of strong water lines. A cooling of the outer atmosphere of a few 100 K, at $\log\tau_{500}<-4$, was needed. This extra outer cooling does, however, not significantly affect the 12.2 $\mu$m HF line, since it is formed deeper in the photosphere: the derived fluorine abundance is only 0.07 dex higher for $\alpha$Boo when using a standard MARCS model compared to using the modified MARCS model of \citet{2002ApJ...580..447R}.

Available chemical evolution models of fluorine in the solar neighborhood predict very similar abundance trends. In Figures \ref{fig:feh}-\ref{fig:koba} we have chosen to compare our results to the chemical evolution models of \citet{2011MNRAS.414.3231K} since those are the only models to our knowledge showing the evolution of fluorine from production in only AGB-stars. For example the models of \citet{2004MNRAS.354..575R} have chemical evolution models of fluorine including (i) the $\nu$-process, (ii) the $\nu$-process and W-R stars, and (iii) $\nu$-process, W-R stars and AGB-stars. Since AGB-stars are the only source of fluorine that has been observationally proven (see Section \ref{sec:intro}) we find the combinations of chemical models in \citet{2011MNRAS.414.3231K} to be more appropriate: they have one including fluorine production only in AGB-stars and two including fluorine production in AGB-stars \emph{and} two different $\nu$-process energies.

From the left panel of Figure \ref{fig:feh}, two observations can be made: first, the models \emph{only} including the AGB star contribution seem to predict the fluorine production compared to iron well within uncertainties, and second, almost all our [F/Fe]-values are slightly sub-solar. The second observation might suggest that the solar value used (A(F)$_{\odot}=4.40$ taken from \citet{2014arXiv1404.5755M}) is too high or that some other process is present. We note, however, that taking the large uncertainty of the solar value ($\sigma A$(F)$=0.25$) into account, our observed trend is in good agreement with the solar value. It would be desirable to determine the solar fluorine abundance to a higher accuracy. However, none of the spectral lines published here or in \citet{2014A&A...564A.122J} below 22 $\mu$m are visible in the photospheric solar atlases of \citet{2010JQSRT.111..521H} and \citet{1994aass.book.....W}. A general estimation of the line strength by $\mathrm{gf} \cdot e^{-\chi_\mathrm{exc}/kT}$ shows that the lines above 22 $\mu$m will only get weaker, meaning that when determining the solar fluorine abundance spectroscopically, one has to use a spectrum from a sunspot. Large uncertainties like the above are then expected because of the uncertainty of the temperature in, and the modeling of, the sunspot.

From the right panel of Figure \ref{fig:feh} we see that $\alpha$Boo (the most metal-poor star in our sample) has an oxygen abundance most consistent with it being of thick disk-type (this is explored in much more detail in \citet{2011ApJ...743..135R}). In the left panel it is indeed slightly better fitted by the thick disk model than the solar neighborhood model, but both models predict higher fluorine than we measure.

In Figure \ref{fig:koba}, we plot [F/O] vs [O/H] to exclude the iron dependence and to better distinguish between contributions from SNIIe and AGB-stars, since iron is abundantly produced in type Ia SNe. We see that the chemical evolution model only including AGB-stars and excluding the $\nu$-process \citep{2011ApJ...739L..57K} best reproduce our fluorine and oxygen abundances. One star, $\alpha$Boo, falls below the line of the solar neighborhood model, but on the other hand, in this case, it is well reproduced by the thick disk model. The uncertainties of our data in Figure \ref{fig:koba} only give room for a $\nu$-process with a neutrino energy much lower than expected \citep{1991NuPhA.527..663H}, which corroborates the results of \citet{2013ApJ...765...51L}.

Based on the combination of Figures \ref{fig:feh}-\ref{fig:koba}, our small sample of stars seem to show that only AGB-stars are needed to explain the fluorine abundance in the solar neighborhood. Eventhough the AGB-star and the $\nu$-process contributions are of similar order in the relatively short metallicity range of our observations \citep{2004MNRAS.354..575R}, including both in the models would most likely over-predict the fluorine abundance. It should, however, be noted that there are certainly uncertainties  in these models, and they depend, to different degrees, on uncertain input values. Therefore, to draw firm conclusions, more observations would be needed: ideally more metal-poor stars to even better test the presence or not of the $\nu$-process, and more metal-rich stars to test the possible production of fluorine in W-R stars. As stated earlier, the 2.3 $\mu$m HF line becomes too weak in metal-poor stars, but the 12.2 $\mu$m line, that is strong in our sample of stars, should be well suited to use for abundance determinations as low as [Fe/H]$\sim -2$. Also further galactic chemical evolution modeling including fluorine production in W-R stars, SNeII and AGB stars together, in different combinations, and independently would be helpful in trying to determine the major contributor of fluorine in the solar neighborhood.

\section{Conclusions}

We present a new line list, with excitation energies and lines strengths, for N-band HF lines and for the first time use one of them for fluorine abundance determination. The abundances derived from this line agrees with the abundances derived from the often used 2.3 $\mu$m line, within uncertainties, for our sample of stars. Thus, our HF lines list for the vibration-rotation lines as presented in \citet{2014A&A...564A.122J} and that for the  pure rotational lines in the N band presented here, give consistent results.

Our measured fluorine-oxygen abundance trend suggests that the fluorine production in AGB-stars might be sufficient to explain the fluorine abundance in the solar neighborhood and that the $\nu$-process is not needed. However, to firmly establish this, more observations are needed. Since the 2.3 $\mu$m line is very weak in metal-poor stars and the N-band lines are much stronger, these lines can probably help determining the possible significance of the $\nu$-process in the chemical evolution of the solar neighborhood. 

\acknowledgments
Kjell Eriksson is thanked for generous help with the modified MARCS model, and the anonymous referee is thanked for insightful comments that helped improve this \textit{Letter} in several ways. This research has been supported by the Royal Physiographic Society in Lund, Stiftelsen Walter Gyllenbergs fond. This research draws upon data as distributed by the NOAO Science Archive. NOAO is operated by the Association of Universities for Research in Astronomy (AURA) under cooperative agreement with the National Science Foundation. This publication made use of the SIMBAD database, operated at CDS, Strasbourg, France, NASA's Astrophysics Data System, and the VALD database, operated at Uppsala University, the Institute of Astronomy RAS in Moscow, and the University of Vienna.

{\it Facilities:} \facility{IRTF (TEXES)}, \facility{Mayall (FTS)}, \facility{ESO:3.6m (HARPS)}, \facility{TBL (NARVAL), \facility{OHP:1.93m (ELODIE)}}.

\end{document}